# Three-dimensional imaging of threading dislocations in GaN by multimodal stimulated Raman scattering microscopy

Shun Takahashi[1], Yusuke Wakamoto[1], Kazuhiro Kuruma[2], Takuya Maeda[1], and Yasuyuki Ozeki[1,2,*]

[1]Department of Electrical Engineering and Information Systems, The University of Tokyo, Bunkyo-ku, Tokyo 113-8656, Japan

[2]Research Center for Advanced Science and Technology, The University of Tokyo, Meguro-ku, Tokyo 153-8904, Japan

*Corresponding author: ozeki@ee.t.u-tokyo.ac.jp

**Abstract**

Emerging gallium nitride (GaN) vertical power devices require high-quality GaN crystals with reduced crystal defects, especially threading dislocations (TDs), to harness the high critical electric field and electron mobility of the material. A strong demand for characterizing TDs has driven the development of many imaging techniques, such as multiphoton excited photoluminescence (MPPL) microscopy and spontaneous Raman microscopy. However, all these techniques lack the capability of visualizing densely existing TDs in a non-destructive and three-dimensional (3D) way or distinguishing the types of TDs. Here, we propose multimodal 3D stimulated Raman scattering (SRS) and MPPL microscopy to non-destructively characterize the TDs in free-standing GaN substrates. Leveraging strong SRS signals, we achieve an imaging speed $10^2$–$10^3$ times faster than that of spontaneous Raman microscopy. This acceleration of the imaging speed enables 3D SRS imaging of the strain fields that are induced by TDs and correlated with the edge components of the Burgers vectors. Multimodal 3D SRS and MPPL imaging reveals rich information on TD features and behavior, including propagation directions, Burgers vectors, dislocation reactions, and their relationships. Furthermore, we present a deterministic identification method for screw-type TDs, which are regarded as killer defects causing leakage current in p-n junctions, by exploiting no SRS peak shifts at screw dislocations in combination with MPPL patterns. Our multimodal imaging technique will greatly advance the understanding of dislocations toward realizing high-quality GaN crystals and high-performance GaN devices.

# Introduction

Gallium nitride (GaN) possesses many superior physical properties, such as a wide band gap, high electron mobility, and high breakdown electric field, making it a unique semiconductor material for various electronic devices. GaN-based high electron mobility transistors (*1*) are widely used in power chargers and wireless base stations. GaN vertical power devices have also been intensively studied to realize high breakdown voltage, low loss, and large current capabilities (*2–4*). However, despite significant crystal quality improvement, the density of threading dislocations (TDs), a kind of crystal defect, in free-standing GaN substrates still reaches $10^4$–$10^7$ cm$^{-2}$. Such a high TD density is a critical issue in emerging vertical power devices because TDs act as leakage current paths in p-n junctions (*5, 6*) and Schottky junctions (*7, 8*), and thus deteriorate device performance and reliability. To make the most of the fascinating material properties, it is essential to establish growth technology for high-quality GaN crystals and better understand the relationship between dislocations and device characteristics. To this end, the GaN community has strongly demanded a three-dimensional (3D) characterization method capable of visualizing densely populated TDs that propagate and react along the growth direction in a growth-condition-dependent manner. Detailed characterization requires information on the Burgers vector (*9*), which is linked to TD types (edge, screw, and mixed types) with different electrical effects (*6, 7*) and governs possible dislocation reactions.

Characterization of TDs in GaN crystals has been demonstrated using various techniques, such as transmission electron microscopy (TEM) (*10, 11*), chemical etching followed by scanning electron microscopy (*12, 13*) or atomic force microscopy (*14, 15*), cathodoluminescence (CL) microscopy (*16*), and X-ray topography (*17*). Nevertheless, none of these imaging techniques fully meet the above-mentioned demands. TEM and chemical etching are the most commonly used but destructive methods, and these two methods as well as CL microscopy are limited to visualizing only thin films (~100 nm) or near the surfaces. Although X-ray topography is a powerful technique capable of fully characterizing sparsely existing dislocations (*17, 18*), a synchrotron radiation facility is required to obtain high-quality images, and its relatively low spatial resolution is unsuitable for typical GaN substrates with a TD density of $10^6$ cm$^{-2}$. To overcome the limitations of the conventional methods, multiphoton excited photoluminescence (MPPL) microscopy (*19, 20*) and micro-Raman spectroscopy (*21–23*) have been applied to the imaging of TDs with deep penetration depth and sub-micrometer spatial resolution. MPPL microscopy was the first to achieve 3D imaging of densely populated TDs over 100 μm in depth along the growth direction (*19*). However, MPPL does not yield information on the Burgers vector. On the other hand, micro-Raman spectroscopy can reveal the edge component of the Burgers vector from a peak shift pattern that appears in the strain field induced by a dislocation (*21*). Its major drawback is the slow imaging speed, which takes several hours to obtain a single image with hundreds by hundreds pixels, making the extension to 3D imaging impractical.

In this work, we propose and demonstrate multimodal 3D stimulated Raman scattering (SRS) microscopy to comprehensively characterize the TDs in free-standing GaN substrates in a non-destructive way. The signal enhancement by SRS (*24–30*) reduces the pixel dwell time to 600 μs, which is a few orders of magnitude shorter than that in spontaneous Raman microscopy. We achieve sensitive detection of the TD-induced strain fields that manifest as SRS peak shift patterns on the order of only 0.01 cm$^{-1}$. Moreover, our SRS microscopy is compatible with MPPL not only in its fast imaging speed but also in its use of near-infrared pulsed light sources and its signal detection scheme. We succeed in simultaneously acquiring SRS and MPPL images that exhibit excellent correspondence and offer the Burgers vectors and TD positions. The 3D mapping of the edge components of the Burgers vectors enables us to track complicated TD behavior including propagation and reactions such as fusion along the crystal growth direction. Furthermore, we demonstrate a deterministic method to identify screw-type TDs, which are particularly important because of the origin of leakage current (*6*, *7*), by combining no SRS peak shift at only screw-type TDs and common MPPL patterns at all types of TDs.

## Results

**Principles of SRS and MPPL to detect TDs**

Both SRS and MPPL are nonlinear optical processes (Fig. 1A). In SRS, synchronized pump and Stokes pulses excite the lattice vibration whose resonance frequency matches the frequency difference of the two pulses, resulting in the power loss of the pump light (stimulated Raman loss, SRL) and gain of the Stokes light (stimulated Raman gain) (*24*, *25*). To excite a specific vibrational mode, narrowband picosecond pulses are employed (*31*). MPPL involves the simultaneous absorption of multiple photons, exciting an electron to the conduction band and leaving a hole in the valence band, and the subsequent recombination of an electron-hole pair releases its excessive energy by emitting a photon (*19*).

Figure 1B illustrates the multimodal SRS and MPPL imaging system. Details on this system are described in Methods. In brief, we use the two-color narrowband picosecond pulses that consist of pump pulses generated by a Ti:sapphire laser at 795 nm and wavelength-tunable Stokes pulses provided by a fiber-based light source (*32*, *33*). The pump and Stokes pulses have the repetition rates of 76 MHz and 38 MHz, respectively, and are synchronized in time by active feedback control. After being combined in space, the pulses are focused onto the sample plane using a water immersion objective with a numerical aperture (NA) of 1.2 to excite the nonlinear processes. SRS signals are obtained by lock-in detection of the transmitted pump pulses, which carry a 38-MHz intensity modulation transferred from the Stokes pulses via SRL, while MPPL signals between 350–400 nm are collected in the epi-detection with a photomultiplier tube. SRS spectral acquisition is executed by wavelength tuning of the Stokes light.

To sensitively detect SRS signals, it is important to suppress competing nonlinear processes (e.g., multiphoton absorption and cross-phase modulation) that disguise themselves as SRL and generate spurious background signals with no chemical specificity (*34*). These background signals increase the fluctuations of the detected signal intensity and thus deteriorate the signal-to-noise ratio (SNR). Our system design addresses this challenge in two ways. First, the use of near-infrared wavelengths (~800 nm) for both pump and Stokes pulses requires more than two photons to exceed the band gap energy of GaN (3.4 eV, 365 nm), decreasing the efficiency of the multiphoton absorption compared to two-photon excitation. Second, the polarization states of the pump and Stokes pulses are adjusted to be perpendicular rather than parallel. While the perpendicular configuration is an unusual choice in biological SRS imaging because of its definitely weaker signal intensity for molecules (*35–37*), the Raman scattering intensities of specific modes in crystals, including the $E_2^H$ mode of GaN, are independent of the polarization configuration (*38*) (Fig. S1). The cross-polarization scheme maintains the SRS signal intensity at the same level while undermining other unwanted nonlinear processes (*39*, *40*) such as multiphoton absorption and cross-phase modulation. In the present system with near-infrared, cross-polarized pump and Stokes pulses, the SRS and MPPL signals are well balanced.

Figure 1C shows the SRS spectrum of a *c*-plane free-standing GaN substrate. The SRS peak of the $E_2^H$ mode is located at 566 cm$^{-1}$ with a full-width at half-maximum (FWHM) of 16 cm$^{-1}$, which is a convolution of the intrinsic Raman spectrum and the spectral widths of the light sources. To precisely extract the peak wavenumber by SRS imaging, we utilize Gaussian fitting for an SRS spectrum measured at nine wavenumbers in the $E_2^H$ mode and at a non-resonant background (Fig. 1D). A Gaussian curve matches the measured SRS spectrum, whose peak shape is mainly contributed by the excitation spectrum. The peak position shifts toward lower or higher wavenumbers depending on whether the applied strain is tensile or compressive. Because the extra half-plane of an edge dislocation extending along the *c*-axis induces an in-plane tensile and compressive strain pair in the *c*-plane (*41*, *42*), as schematically depicted in Fig. 1E, edge- and mixed-type TDs with the half-plane can be detected by SRS imaging. The SRS peak shift pattern corresponding to the strain field reveals not only the TD position but also the edge component of the Burgers vector $\mathbf{b}_{\text{edge}}$.

**Simultaneous SRS and MPPL imaging**
We performed SRS and MPPL imaging of a free-standing *c*-plane *n*-type GaN substrate with a carrier concentration of $10^{17}$ cm$^{-3}$. Both in SRS and MPPL imaging, the pixel dwell time was 4 µs, and the number of integrations was 15. Figure 2A shows the acquired MPPL image with numerous dark spots formed around TDs, which act as non-radiative recombination centers (*19*). The number of the dark spots is 104 within the area of 80 × 80 µm$^2$, corresponding to a TD density of $1.6 \times 10^6$ cm$^{-2}$, which is a typical value for commercial free-standing GaN substrates. The diameter of these spots is determined by the convolution of the focal spot area of excitation pulses and the diffusion length of minority carriers (*43*). The signal intensities of MPPL involving either pump or

Stokes photons are proportional to $P_p^{3.31}$ and $P_S^{3.58}$ (Fig. S2), where $P_p$ and $P_S$ are the pump and Stokes powers, respectively. These exponent values (3.31 and 3.58) indicate that the MPPL process requires mainly three or four photons for excitation. In addition, MPPL signals are efficiently produced at higher peak powers that are provided when the pump and Stokes pulses overlap (Fig. S3).

Figure 2B shows the corresponding SRS peak shift image after image processing of SRS spectral data. Averaging and high-pass filtering effectively remove noise and extract the SRS peak shift caused by TDs. Details on the image processing can be found in Methods. There are many SRS patterns featuring a pair of positive and negative peak shifts (Fig. 2, B to D), which appear in the strain field around edge- and mixed-type TDs. The good agreement between the positions of the peak shift patterns (Fig. 2B) and the dark spots (Fig. 2A) validates that our SRS microscopy can accurately detect TDs. Considering the SRS spectral imaging at ten wavenumbers, the accumulation time per pixel is calculated to be 600 μs (= 4 μs × 15 integrations × 10 wavenumbers). This effective pixel dwell time is $10^2$–$10^3$ times smaller than that previously required to achieve a sufficient SNR in spontaneous Raman microscopy (~0.2–1 s) (*21*, *23*).

We analyzed SRS images to quantify the amount of peak shift (Fig. 2, E and F) and examine the direction of $\mathbf{b}_{edge}$ (Fig. 2G). To quantify the peak shift, the SRS image shown in Fig. 2B was divided into 104 sub-images centered on individual TDs and the remaining background region. The standard deviation of the peak shift in the background region is as small as 0.005 cm$^{-1}$ (Fig. 2E), and the maximum positive and negative peak shifts in each of the 104 sub-images are distributed around ±0.03 cm$^{-1}$ (Fig. 2F). The sensitivity to peak shifts is determined by the spectral width and SNR (see Discussion for details). Despite its modest spectral resolution, the SRS imaging system with a high SNR enables us to sensitively detect the TD-induced peak shift on the order of 0.01 cm$^{-1}$ under the small background fluctuation. Next, to evaluate the $\mathbf{b}_{edge}$ direction, we collected a total of 13 SRS peak shift images including Fig. 2B from different lateral locations. The direction of $\mathbf{b}_{edge}$ was determined by calculating the normalized cross-correlations between an SRS peak shift pattern around a TD and 36 template patterns. These templates sharing the same basic structure are rotated by 10-degree increments. The histogram (Fig. 2G) shows the $\mathbf{b}_{edge}$ directions of 888 isolated TDs excluding those which are located close to other TDs or have low correlation with all of the template patterns. The distribution is clearly concentrated near the *a*-axes (⟨11$\bar{2}$0⟩) of the GaN substrate in accordance with hexagonal crystal symmetry (*9*). Thus, our SRS microscopy can reveal the direction of $\mathbf{b}_{edge}$ by analyzing the SRS pattern with positive and negative peak shifts.

Simultaneous MPPL and SRS imaging provides the information on the TD position (circle markers) and the $\mathbf{b}_{edge}$ direction (arrows), respectively (Fig. 2H). The TD positions are tracked using TrackMate (*44*), a Fiji plugin, and the $\mathbf{b}_{edge}$ directions are specified based on the normalized cross-correlations under the premise (*9*) verified in Fig. 2G that basis vectors of $\mathbf{b}_{edge}$ are restricted along the *a*-axes. TDs without assigned arrows indicate that it is difficult to determine the Burgers vector due to the weak peak shift of a largely inclined dislocation (elongated dark spots) or the peak shift pattern cancelled out by overlapping TDs (darker spots). A higher sensitivity would allow such ambiguous peak shift patterns to be clarified. The current sensitivity and spatial resolution of our SRS imaging system are sufficiently high to resolve complex strain fields generated by multiple TDs close to each other (Fig. 2I). Unlike the size of MPPL dark spots, that of SRS peak shift patterns is independent of the carrier density, which benefits diffraction-limited SRS imaging of TDs. SRS microscopy can achieve both high spatial resolution and signal intensity owing to its nonlinear nature, whereas spontaneous Raman microscopy is subject to a trade-off between them. The SRS images shown in Fig. 2I visualize the strain fields created by two TDs with distances less than 1 μm, and agree well with the template patterns that are superpositions of basic template patterns considering a single isolated TD. The number of possible template patterns for the two closely positioned TDs is only 36 ($=6^2$), because the center positions of the two TDs are known from the MPPL image, and the $\mathbf{b}_{edge}$ direction is restricted to the six directions. Multimodal imaging allows the unique determination of the $\mathbf{b}_{edge}$ directions of multiple TDs from their positions and SRS peak shift pattern.

**Multimodal 3D SRS and MPPL imaging**

The fast imaging speed and deep penetration depth of our SRS microscopy enable its extension to three dimensions, providing a powerful tool for investigating TDs along the growth direction of GaN crystals. We acquired a multimodal 3D image set by obtaining SRS and MPPL images over a 55 μm range with a 2.5 μm step along the *c*-axis. The 3D image set contains the images shown in Fig. 2, A and B. Figure 3A shows the full plot of the multimodal 3D imaging. The growth direction of the GaN substrate processed by hydride vapor phase epitaxy (HVPE) is along the *c*-axis ([0001]), that is, from the bottom to the top in the 3D plot. The positions of TDs (circle markers) and the directions of the Burgers vectors (colors) are derived from the MPPL and SRS images, respectively. Most of the TDs propagate parallel to the *c*-axis, while some dislocation lines are largely inclined from the *c*-axis, with angles sometimes reaching 45 degrees (Fig. S4). The projection of these inclined dislocation lines onto the *c*-plane is aligned with the *a*-axes (Fig. S4A). The directions of $\mathbf{b}_{edge}$ are maintained during propagation when no reactions with other TDs occur. The whole set of SRS and MPPL images can be found in Movie S1. Many fusion/splitting reactions of TDs occur within the measured area, whereas no annihilation reactions are observed. Closely positioned TDs, indicated as TDs 1–5 in Fig. S4A, propagate keeping their relative positions constant. This positional stability of the TDs whose strain fields overlap is probably related to the minimization of the total dislocation energy proportional to $|\mathbf{b}|^2$ (*45*). Moreover, among those TDs, only TDs 1 with opposite $\mathbf{b}_{edge}$ directions stay at the same position in the lateral direction.

Figure 3B shows the 3D plot of the selected area and the corresponding TD mapping, MPPL, and SRS images. The attached numbers in the TD mapping images are manually assigned so that each TD can be easily identified. TD 1, 5a, and 6 have largely inclined dislocation lines, which are a feature of screw-type TDs (*20*), and the directions of their in-plane projections are parallel to the directions of $\mathbf{b}_{edge}$. The relationship between an inclined dislocation line and the Burgers vector will be useful to classify the TD types in stressed layers, where it is parallel for mixed-type TDs and perpendicular for edge-type TDs (*22*, *46*). Multimodal 3D imaging can also visualize interactions of multiple TDs. For instance, TD 5a begins to move substantially in the lateral direction when TD 5a is approached by TD 6 between $Z = 25$ µm and $Z = 12.5$ µm. In addition, TD 5 splits into TD 5a and 5b between $Z = 35$ µm and $Z = 25$ µm while keeping the sum of the $\mathbf{b}_{edge}$ direction from a purple marker to blue and red markers. TD 1 and 2 also maintain their $\mathbf{b}_{edge}$ directions (and probably TD 3 and 4 as well). The conservation law of the Burgers vector has been theoretically constructed (*9*, *45*) and experimentally observed using other methods (*47*), but these results presented here are the first 3D verification of this principle within a densely TD-populated bulk GaN crystal. Our 3D SRS and MPPL imaging provides rich information on TD properties such as propagation directions, Burgers vectors, and their relations to comprehensively characterize TDs.

**Identification of screw dislocations**

The origin of reverse leakage current caused by TDs in GaN power devices have been intensively studied from diverse perspectives, such as TD types (edge, screw, and mixed) (*6*), dislocation core structures (*48*), diffusion of impurities (*47*, *49*), and forward current stress (*50*). Although its physical mechanism remains under debate, an important consensus is that screw-type TDs form leakage current paths in p-n junctions (*6*, *48*, *50*), making their differentiation from other types of TDs essential for realizing high-performance vertical power devices. However, it is challenging to reliably identify and track screw-type TDs in a non-destructive way. Type-dependent statistical trends in dislocation lines and dark spot contrasts were reported in MPPL microscopy (*20*), but MPPL imaging alone is insufficient for the deterministic identification of screw-type TDs. We demonstrate that multimodal SRS and MPPL imaging can deterministically identify screw-type TDs by exploiting the absence of SRS peak shift patterns at screw dislocations, in contrast to MPPL dark spots that appear at both edge and screw dislocations. No or too small SRS peak shift at screw dislocations stems from its insensitiveness to shear strain (*21*, *23*, *51*, *52*). Screw-type TDs exclusively induce shear strain without tensile or compressive strain.

The sample is another free-standing *c*-plane *n*-type GaN substrate with a carrier concentration of >5 × $10^{18}$ cm$^{-3}$. Instead of referring to the classification by TEM or an etch pit technique, we focus on the area enclosed by the green square in the MPPL image (Fig. 4A). The dark spot at the center has two conspicuous features: bright six-fold symmetry lines and a high contrast compared to the other dark spots. These features of the dark spot are consistent with those attributed to two closely (~40 nm (*53*) or ~100 nm (*54*)) spaced mixed-type TDs with opposite edge components of their Burgers vectors in the literature (*53*, *54*). The TDs are not pure screw-type TDs but, like them, have a combined Burgers vector with only the screw component. Figure 4B shows the SRS and MPPL images obtained in the area of interest at depths between 3.5 µm and 13.5 µm from the Ga-polar surface. Only the central dark spot indicated by the black arrow produces no detectable peak shift in the SRS images, whereas the other dark spots induce familiar peak shift patterns with a positive and negative pair. The absence of the SRS peak shift pattern corresponds to $\mathbf{b}_{edge}$ equal to zero. Thus, using our multimodal approach, screw-type TDs can be deterministically identified by comparing their SRS and MPPL patterns.

**Discussion**

We have demonstrated multimodal 3D SRS and MPPL microscopy to comprehensively characterize the TDs in GaN substrates. Our SRS imaging is capable of detecting the SRS peak shift as small as 0.03 cm$^{-1}$ with an effective pixel dwell time of 600 μs. This fast imaging speed of SRS enables us to achieve 3D imaging of TDs in combination with MPPL. Multimodal 3D imaging reveals rich information on TDs such as positions, propagation directions, reactions, Burgers vectors, and relationships between these features/events. Furthermore, we demonstrate the effectiveness of our multimodal approach to deterministically identify screw-type TDs. This powerful imaging technique will facilitate detailed dislocation characterization toward high-quality GaN crystals and provide non-destructive wafer inspection in industrial applications.

Our SRS imaging still has room for improvement in sensitivity to peak shifts. We evaluated the SNR of SRS signal detection in our imaging system (Fig. S5) and simulated the sensitivity to peak shifts as a function of SNR (Fig. S6). As shown in Fig. S5A, the SRS signal intensity in the $E_2^H$ mode increases linearly with the Stokes power. At high signal levels, the noise defined by the standard deviation of SRS signal intensities also increases proportionally to the SRS signal (Fig. S5B), resulting in a saturated SNR (Fig. S5C). We attribute the sources of the signal-intensity-dependent noise to the intensity noise of the Stokes laser, shift in the excitation wavenumber, and timing jitter of pulse synchronization. This situation is completely different from SRS imaging in biological applications (*55*), where only the noise of the light source to be detected is of concern because it is the dominant factor to determine the noise level in signal detection. Suppressing those excess noise factors by achieving a more stable and quiet Stokes light source and tighter pulse synchronization could reduce the noise to the shot-noise level, where the SNR scales without limitation as the pump and Stokes powers increase. Figure S6 shows the relationship between the SNR and sensitivity to peak shifts in three spectral FWHMs. Assuming that the SNR of our imaging system is ~50 with a pixel dwell time of 4 μs (see Fig. S5C), the root mean square error of the SRS peak shift shown in Fig. 2E (i.e., $\sigma = 0.005$ cm$^{-1}$) is reasonable after SNR enhancement by integrating 15 frames and averaging 5 × 5 pixels. The simulation results indicate that we could achieve more sensitive and faster SRS imaging by a higher SNR or narrower excitation spectrum.

## Methods

### Imaging system

Our multimodal imaging system is based on the previously developed light source and SRS imaging system (*32, 33*). The pump pulses generated by a Ti:sapphire laser (Mira900D, Coherent) have a repetition rate of 76 MHz at 795 nm. The Stokes light source is a widely wavelength-tunable fiber optical parametric oscillator (FOPO) (*32, 33*) pumped by a Yb-doped fiber laser with a repetition rate of 38 MHz, a half of 76 MHz. A main fiber amplifier in the fiber laser employs a single-clad gain fiber and two single-mode laser diodes (LDs) in the present system, rather than a double-cladding large-mode-area gain fiber and a multimode LD in the previous system (*33*), to stabilize the excitation of the FOPO. The pump and Stokes pulses are synchronized by monitoring their two-photon absorption signal and applying the active feedback to a phase modulator and piezo stage in the fiber laser (*56*). The polarization states of linearly polarized pump and Stokes pulses are perpendicular. The two pulses are combined at a dichroic mirror (DMSP805R, Thorlabs), scanned by a dual-axis galvanometer scanner (GVS002, Thorlabs), and focused onto the sample plane using a water immersion objective (60x, NA 1.2, UPLSAPO60XW, Olympus). The pump pulses collected by another objective (UPLSAPO60XW, Olympus) are transmitted through a short-pass filter (FESH0800, Thorlabs) with the Stokes pulses blocked, and are incident on a photodiode whose output is connected to a home-made lock-in detector circuit and a data acquisition (DAQ) system (USB-6363, National Instruments). Photons radiated by the near-band-edge emission are reflected by a dichroic mirror (DMLP650R, Thorlabs) below the first objective and transmitted through a short-pass filter (#84-712, Edmund Optics) for blocking excitation pulses and a band-pass filter (#12-096, Edmund Optics) for limiting the wavelength range of detected PL signals to 350–400 nm. The MPPL signals are epi-detected with a photomultiplier tube (H10723-210, Hamamatsu Photonics) connected to the DAQ system. The sample stage equipped with a DC servo actuator automatically moves in the axial direction.

### Imaging conditions

Imaging conditions are summarized here. We acquired all SRS and MPPL images with a pixel dwell time of 4 μs and 15 integrations. The images shown in Fig. 2, A and B, Fig. 4A, and Movie S1 have a field of view (FOV) of 80 × 80 μm$^2$ and a size of 512 × 512 pixels, and those shown in Fig. 4B have an FOV of 30 × 30 μm$^2$ and a size of 256 × 256 pixels. The pump and Stokes powers on the sample plane were approximately 22 mW and 21 mW, respectively. It was possible for the light sources to increase the pulse powers without sample damage, but higher powers were less effective in improving the SNR, as described in Discussion. The discrepancy between the axial movement of the sample stage and the actual displacement of the focal spot due to the refractive index mismatch (GaN: 2.35, water: 1.33, at 800 nm) was corrected.

**GaN substrate samples**

We used two commercially available, *c*-plane free-standing *n*-type GaN substrates grown by HVPE. The first GaN substrate, used for the experiments in Figs. 1 to 3, has a carrier concentration of $10^{17}$ cm$^{-3}$ and a thickness of 350 μm. The other substrate from a different supplier was used for the screw dislocation identification experiment (Fig. 4). It has a high carrier concentration of $>5 \times 10^{18}$ cm$^{-3}$ by germanium doping and a thickness of 400 μm. The substrates were placed on 150-μm thick coverslips with the Ga-polar surface facing down. One of the *a*-axes of the GaN substrates was aligned with the vertical axis of the images.

**Image processing**

A five-step image processing pipeline was applied to convert the raw SRS spectral data into an SRS peak shift image. We first integrated each set of 15 SRS images that were continuously acquired at a specific wavenumber, and then applied an averaging filter with a window size of 5 × 5 pixels to each SRS image. The image integration and spatial averaging reduced low- and high-frequency noise, respectively. Next, background signals acquired at a non-resonant wavenumber were subtracted from the SRS signals in the $E_2^H$ mode to fix the spectral baseline at zero in the following fitting process. We derived the peak wavenumber by pixel-wise fitting using a Gaussian curve with three parameters: the peak wavenumber, intensity, and width. The resulting SRS peak wavenumber image (Fig. S7) contains two distinct background patterns: fine horizontal lines along the fast-scan axis and global non-uniformity. Both patterns are artifacts due to system fluctuations and inhomogeneity, not strain fields in the substrate. Fortunately, the latter pattern is composed of a spatial frequency much lower than that of the TD-induced peak shift pattern. Exploiting this frequency difference, we removed the global non-uniformity by calculating the peak shift relative to the locally averaged peak wavenumber at each pixel. The final processing step yielded a high-pass filtered SRS peak shift image (Fig. 2B). The pixel value of the image represents the average SRS peak shift within the excitation volume convolved with the spatial filtering area.

**Acknowledgements**

We would like to thank Yoshitaka Shirasaki and Zhuohao Yang for their helpful discussions, and Riena Jinno for technical support. This work was supported by Japan Society for the Promotion of Science (JSPS) KAKENHI (JP23H00271, JP23KJ0729, JP25K22085), Japan Science and Technology Agency (JST) Core Research for Evolutional Science and Technology (CREST) (JPMJCR2331), and Ministry of Education, Culture, Sports, Science and Technology (MEXT) (JPMXS0118067246).

**Author contributions**

S.T., T.M, and Y.O. conceived the idea for the project. S.T. constructed the light source and imaging system. S.T. performed the experiments and analyzed the data with help from Y.W. and K.K. All authors discussed the results. T.M. and Y.O. supervised the project. S.T. and Y.O. wrote the manuscript with input from all authors.

**Competing interests**

The authors declare no competing interests.

**Data availability**

The data presented in this work are available from the corresponding author on reasonable request.

**Supplementary Materials**

Figs. S1 to S7

Movie S1

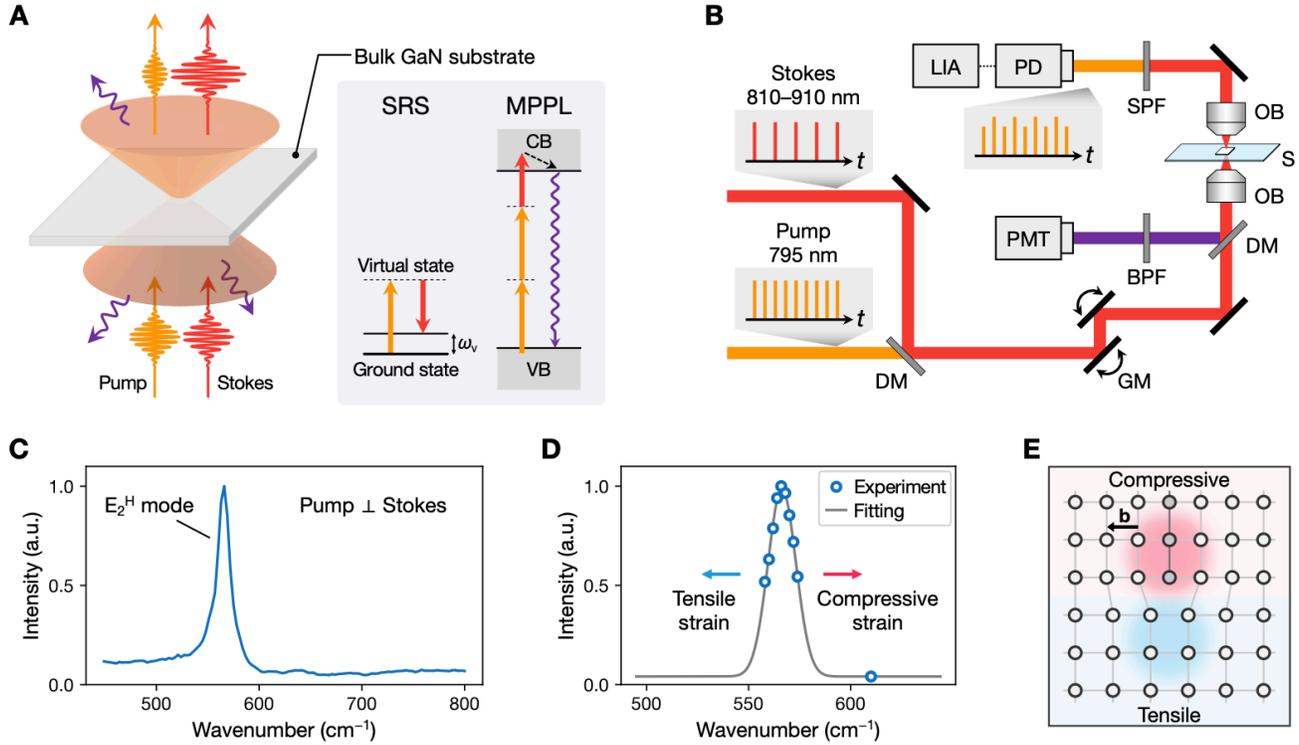

**Fig. 1. Principles of SRS and MPPL to detect TDs.** (**A**) Energy diagram of SRS and MPPL. In SRS, two-color synchronized pump and Stokes pulses excite the frequency-matched lattice vibration and experience the power loss in pump (stimulated Raman loss) and gain in Stokes (stimulated Raman gain). In MPPL, electrons are excited from the valence band to the conduction band through multiphoton absorption, and their recombination emits photons with a photon energy corresponding to the band gap of GaN (3.4 eV, 365 nm), which slightly widens with higher electron concentrations. VB, valence band; CB, conduction band. (**B**) Schematic of the multimodal imaging system. The pump wavelength is fixed at 795 nm, and the Stokes wavelength (~832 nm) is widely tunable to excite the $E_2^H$ mode of GaN and obtain SRS spectra. The polarization states of the pump and Stokes pulses are perpendicular to suppress unwanted nonlinear processes competing with SRS. The SRS and MPPL signals are collected in transmission and reflection geometries, respectively. DM, dichroic mirror; GM, galvanometer scanner; OB, objective; S, sample; SPF, short-pass filter; BPF, band-pass filter; PD, photodiode; LIA, lock-in amplifier; PMT, photomultiplier tube. (**C**) SRS spectrum of a *c*-plane free-standing GaN substrate. Only the $E_2^H$ mode is observed in the cross-polarized pump–Stokes configuration. (**D**) Derivation of the SRS peak wavenumber. The peak wavenumber is derived using Gaussian fitting of the SRS signals obtained at nine wavenumbers in the $E_2^H$ mode and at a non-resonant wavenumber. It shifts toward lower or higher wavenumbers depending on tensile or compressive strain. (**E**) Image diagram of an edge dislocation. Its inserted extra half-plane induces an in-plane tensile and compressive strain pair around the center of the dislocation. The black arrow denotes the Burgers vector.

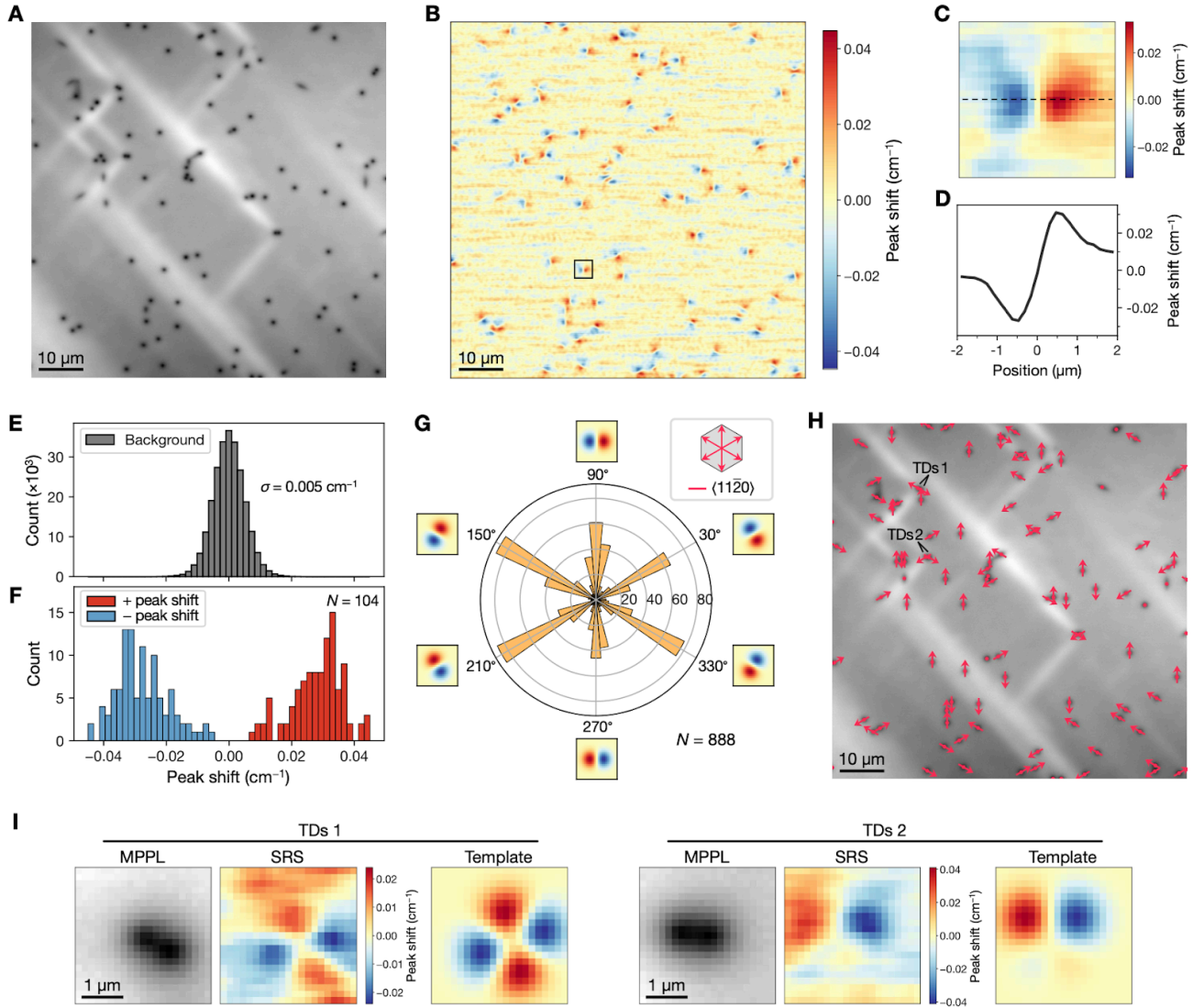

**Fig. 2. Simultaneous SRS and MPPL imaging and image evaluations.** (**A**) MPPL image of the TDs in a free-standing GaN substrate. Dark spots indicate the locations of TDs, which act as non-radiative recombination centers. The number of dark spots is 104 within the area of 80 × 80 μm$^2$, corresponding to a TD density of 1.6 × 10$^6$ cm$^{-2}$. (**B**) SRS image of the same area as (A). The SRS peak shift patterns with a positive and negative pair visualize the strain fields induced by edge- and mixed-type TDs. The positions of the peak shift patterns agree well with those of the dark spots in the MPPL image. (**C**) Enlarged image of the SRS peak shift pattern indicated by the black square box in (B). (**D**) Peak shift profile along the dashed line in (C). (**E** and **F**) Histograms of the peak shift in the background (E) and around TDs (F) in the SRS image (B). In the background, the standard deviation of the peak shift is 0.005 cm$^{-1}$. Around TDs, the values defined by the maximum positive and negative peak shifts of each TD-induced strain field are distributed around ±0.03 cm$^{-1}$. (**G**) Histogram counting the direction of the edge component of the Burgers vector **b**$_{edge}$. A total of 888 peak shift patterns in 13 SRS images are collected to statistically evaluate the direction. The **b**$_{edge}$ directions are clearly distributed near the *a*-axes (⟨11$\bar{2}$0⟩) of the GaN crystal. (**H**) MPPL image with markers of the TD positions (circles) and six **b**$_{edge}$ directions (arrows), which are derived from the MPPL and SRS images, respectively. (**I**) Enlarged MPPL and SRS images of closely positioned TDs indicated in (H). The template patterns are superpositions of two peak shift patterns with a positive and negative pair. The SRS images visualize the complicated strain fields created by two TDs and show good agreement with the template patterns.

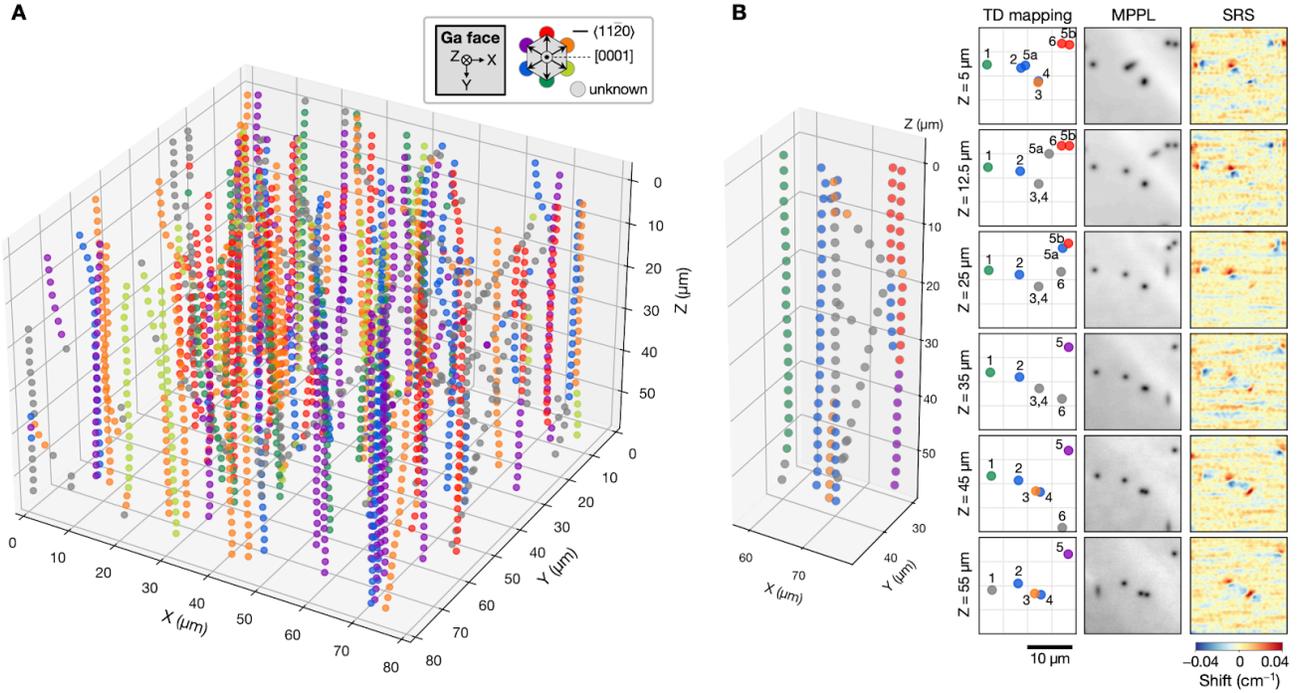

**Fig. 3. Multimodal 3D SRS and MPPL imaging.** (**A**) Full 3D mapping of the edge component of the Burgers vector $\mathbf{b}_{edge}$. The circle marker and its color represent the TD position and the $\mathbf{b}_{edge}$ direction, respectively. The 3D data consist of SRS and MPPL images that are sequentially acquired along the $c$-axis with a range of 55 µm and a step of 2.5 µm. The crystal growth direction of the substrate processed by HVPE is from bottom to top in the 3D coordinates. (**B**) TD behavior within the area of interest in (A) and corresponding TD mapping, MPPL, and SRS images at several $z$ positions. The identification numbers in the TD mapping images are manually assigned to distinguish and track each TD. The SRS colormap is clipped at −0.04 and 0.04 for visibility.

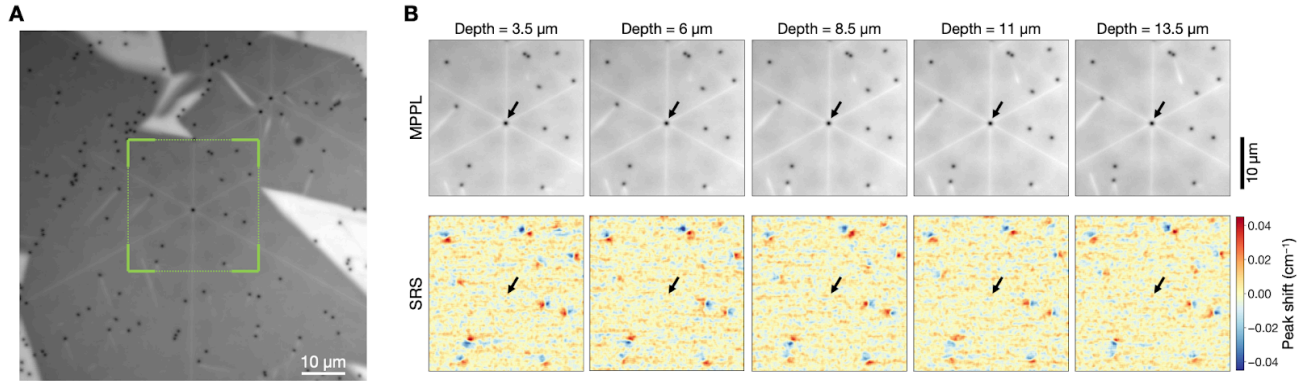

**Fig. 4. Identification of screw dislocations.** (**A**) MPPL image of a different free-standing GaN substrate. The diameter of dark spots is smaller than that in the previous GaN substrate due to the shorter diffusion length of minority carriers in this highly doped *n*-type GaN substrate. (**B**) SRS and MPPL images within the area indicated in (A). The dark spot at the center of the MPPL images has two features: bright six-fold symmetry lines and a higher contrast than the other dark spots. These features are consistent with those of two very close mixed-type TDs whose opposite edge components of the Burgers vectors are cancelled out. In the SRS images, only the dark spot at the center produces no peak shift pattern, whereas the other dark spots exhibit the peak shift patterns. Utilizing the absence of SRS peak shifts at screw dislocations, multimodal SRS and MPPL imaging can deterministically identify screw-type TDs. Note that MPPL images alone are insufficient to identify screw-type TDs because not all screw-type TDs have noticeable features in the MPPL images. The SRS colormap is clipped at −0.045 and 0.045 for visibility.

# Supplementary Material for:

# Three-dimensional imaging of threading dislocations in GaN by multimodal stimulated Raman scattering microscopy

Shun Takahashi,[1] Yusuke Wakamoto,[1] Kazuhiro Kuruma,[2] Takuya Maeda,[1] and Yasuyuki Ozeki[1,2,*]

[1]Department of Electrical Engineering and Information Systems, The University of Tokyo, Bunkyo-ku, Tokyo 113-8656, Japan

[2]Research Center for Advanced Science and Technology, The University of Tokyo, Meguro-ku, Tokyo 153-8904, Japan

[*]Corresponding author: ozeki@ee.t.u-tokyo.ac.jp

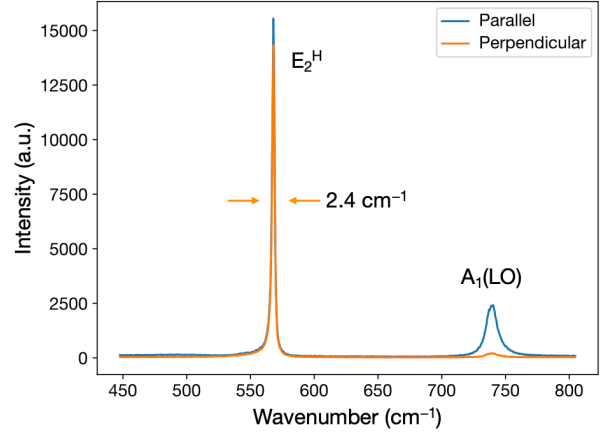

**Fig. S1. Spontaneous Raman spectra of GaN in two scattering configurations.** Using a high-precision micro-Raman spectrometer (LabRAM Odyssey, Horiba) with a CW laser at 532 nm, we measured the spontaneous Raman spectra of a free-standing GaN substrate in the $z(x,x)\bar{z}$ (parallel) and $z(x,y)\bar{z}$ (perpendicular) configurations, where the z-axis is parallel to the c-axis, the x- and y-axes lie in the c-plane, and the first/last and second/third characters represent the propagation and polarization directions of the incident/scattered light, respectively. The perpendicular configuration in spontaneous Raman measurements corresponds to the cross-polarized pump and Stokes configuration in SRS measurements. The Raman peak of the $E_2^H$ mode exists at 568 cm$^{-1}$ with a spectral FWHM of 2.4 cm$^{-1}$. Although there is a slight difference probably due to the polarization dependence of the measurement setup, the $E_2^H$ peak intensities of the two spectra are almost the same.

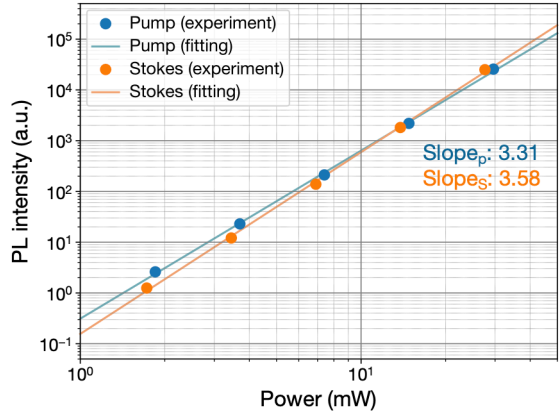

**Fig. S2. MPPL signal intensity as a function of input power.** Only the pump (795 nm, 1.56 eV) or Stokes (832 nm, 1.49 eV) beam was incident on the sample with the other beam blocked. In the logarithmic plot, the slopes of the linear fitting lines are 3.31 for the pump and 3.58 for the Stokes. These values indicate that this MPPL mainly follows three- and four-photon excitation processes, including multiphoton absorption exceeding the band gap of GaN (3.4 eV) and transitions via defect levels.

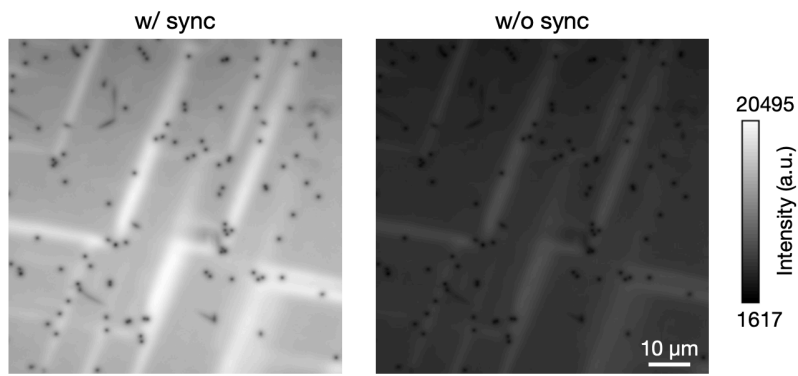

**Fig. S3. MPPL images with (w/) and without (w/o) synchronization of the pump and Stokes pulses.** Without pulse synchronization, the two pulses have slightly different repetition rates and rarely overlap in time. In this case, multiphoton excitation processes involve either pump or Stokes photons. The much higher signal level with synchronization benefits from the higher peak power when the pump and Stokes pulses overlap, making the nonlinear processes efficient.

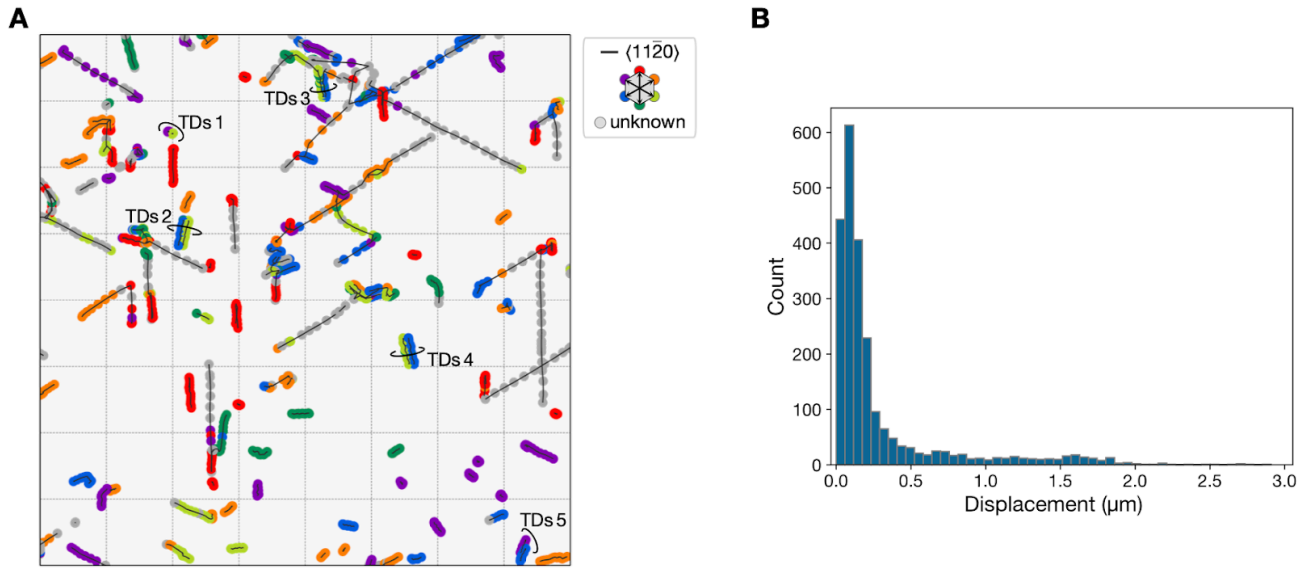

**Fig. S4. Tracking of the TD displacement.** (**A**) Projection of the 3D plot of $\mathbf{b}_{edge}$ (Fig. 3A) onto the *c*-plane. The marker colors and lines represent the $\mathbf{b}_{edge}$ directions of TDs and TD displacements between adjacent images, respectively. Tracking of the TD displacement was performed with TrackMate, a Fiji plugin. The dashed lines are spaced 10 μm apart, and the black arrows in the upper right corner represent the *a*-axis directions. The majority of TDs stay almost at the same position in the lateral direction, whereas others move significantly. Even for the same TD, its displacement varies in the middle, which results in a switch between detectable and undetectable $\mathbf{b}_{edge}$ directions. TDs 1–5 are TD pairs that consist of two closely positioned TDs with stable relative positions during propagation. (**B**) Histogram of the lateral TD displacement between two sequential images 2.5 μm apart in the axial direction along the *c*-axis.

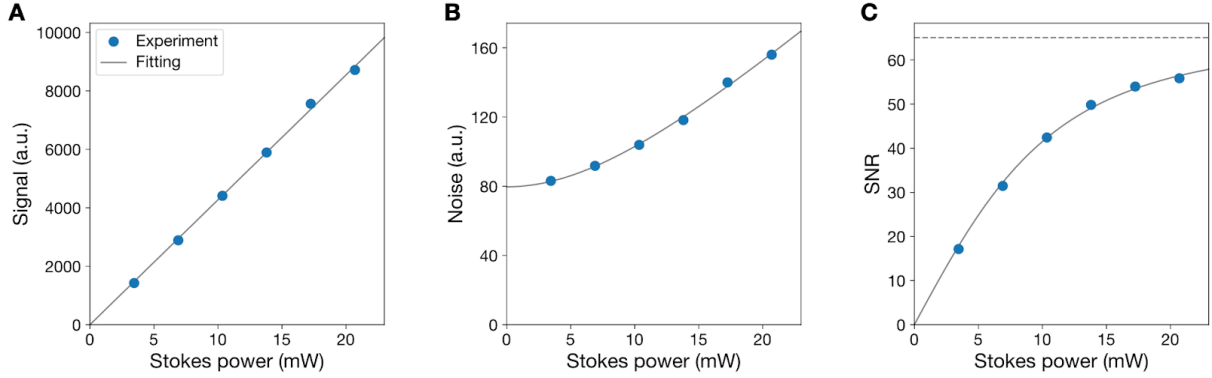

**Fig S5. SNR of SRS signal detection.** (**A** to **C**) The signal intensity (A), noise (B), and SNR (C) were measured at the $E_2^H$ peak of GaN as a function of Stokes power $P_S$. The pump power was set to 14.8 mW. SRS signals were detected 10,000 times with a dwell time of 4 µs. (**A**) The average signal intensity ($S$) is linearly proportional to the Stokes power. The fitting line is defined by the simple relation $S = c_1 P_S$, where $c_1$ is a coefficient. (**B**) The noise ($N$) is the standard deviation of the SRS signal intensities. This definition of noise differs from that in biological SRS imaging, where noise is usually defined as the standard deviation of background signals at a non-resonant wavenumber. The experimental results (blue circles) agree well with the fitting curve (gray line) following the equation $N = (c_2 P_S^2 + c_3)^{1/2} = (c_4 S^2 + c_3)^{1/2}$, where $c_2$, $c_3$, and $c_4$ are coefficients, and $c_3$ corresponds to the noise independent from the SRS signal level, such as the circuit noise and the shot noise of the pump laser. The noise level increases in proportion to the Stokes power (i.e., the signal intensity $S$), meaning that excess noise is dominant at high signal levels. This excess noise is mainly attributed to fluctuations in the Stokes power, excitation wavenumber, and pulse synchronization. (**C**) SNR (= $S/N$) increases to 55.9 at the Stokes power of 20.7 mW. The gray dashed line represents the saturated SNR.

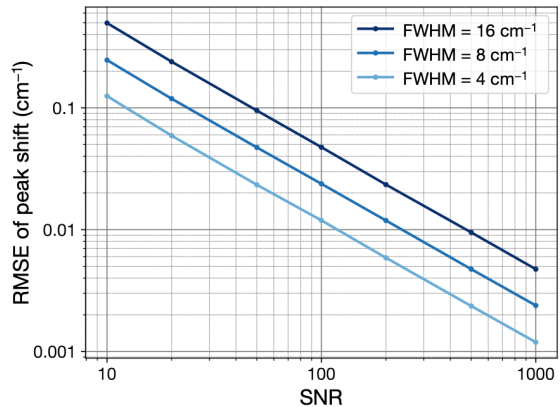

**Fig. S6. Relationship between SNR and sensitivity to peak shifts.** Using numerical analysis (*1*), we simulated the root mean square error (RMSE) of the peak wavenumber as a function of SNR. The SNR is defined in the same way as in Fig. S5. The simulation procedure is as follows. First, we prepared an artificial spectrum that mimicked the experimentally obtained SRS spectrum, by creating a Gaussian profile with a certain peak value $I_{peak}$ and FWHM, and then sampled its signal values at nine equally spaced wavenumbers within the FWHM (i.e., $\Omega_{sampling}$ = −FWHM/2, …, 0, …, FWHM/2). Normally distributed noise with a standard deviation of $I_{peak}$/SNR was added to the signal. Second, we calculated the peak wavenumber $\Omega_{peak}$ of the prepared spectra by Gaussian fitting and derived the RMSE in 10,000 iterations. The figure shows that the RMSE is inversely proportional to the SNR and proportional to the spectral FWHM. Therefore, a higher SNR and narrower spectrum lead to a higher sensitivity to SRS peak shifts.

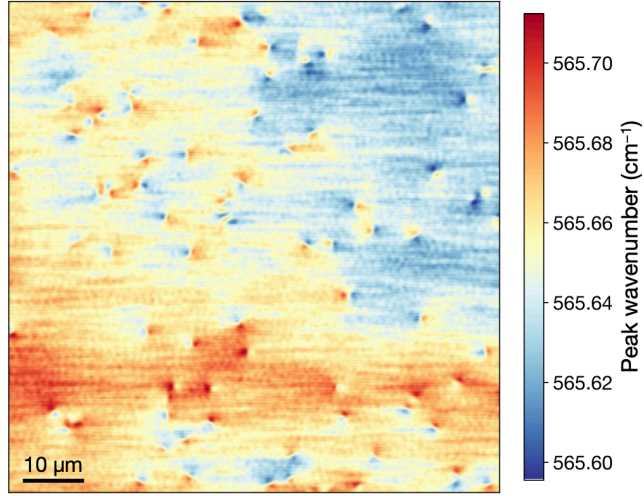

**Fig. S7. SRS image of the peak wavenumber.** Characteristic small SRS patterns with a wavenumber up- and down-shift pair reflect the compressive and tensile strains around the TDs. Meanwhile, uneven peak wavenumbers are observed in the background. These background patterns are attributed to system inhomogeneity and fluctuations, such as in the laser power and excitation wavenumber. The overall range of the peak wavenumber is less than 0.12 cm$^{-1}$.

**Movie S1. 3D SRS and MPPL imaging.** Three images are placed in a row. From left to right, the images are the MPPL image, SRS image, and TD mapping image with indications of the TD positions derived from the MPPL image and the edge components of the Burgers vectors derived from the SRS image. From the first to the last images, the $z$ position changes along the $c$-axis direction (i.e., the crystal growth direction) with a range of 55 µm and a step of 2.5 µm. The SRS colormap is clipped at −0.045 and 0.045 for visibility. Scale bars, 10 µm.